\input harvmac
\input epsf

\def\R{\relax{\rm I\kern-.18em R}}
\font\cmss=cmss10 \font\cmsss=cmss10 at 7pt
\def\Z{\relax\ifmmode\mathchoice
{\hbox{\cmss Z\kern-.4em Z}}{\hbox{\cmss Z\kern-.4em Z}}
{\lower.9pt\hbox{\cmsss Z\kern-.4em Z}}
{\lower1.2pt\hbox{\cmsss Z\kern-.4em Z}}\else{\cmss Z\kern-.4em
Z}\fi}\
\def\np{Nucl. Phys. }
\def\pl{Phys. Lett. }
\def\pr{Phys. Rev. }

\def\prl{Phys. Rev. Lett. }

\def\sinh{{\rm sinh}}
\def \cosh{{\rm cosh}}

\def\Tr{{\rm Tr}}

\def\CN{{\cal N}}

\def\CN0{{\cal N}_0}


\Title{}
{\vbox{
\centerline{Anomalies and large $N$ limits in matrix string theory}
}}
\lref\bsfs{T.~Banks, W.~Fischler, S.H.~Shenker and L.~Susskind, \pr
{\bf D55}(1996) 5112-5128, hep-th/9610043.}
\lref\banks{T. Banks, \np Proc. Suppl. {\bf 67} 180-224 (1998), 
hep-th/9710231.}
\lref\BS{D. Bigatti and L. Susskind, hep-th/9712072.}
\lref\Tayl{W. Taylor, hep-th/9801021.}
\lref\suss{L. Susskind, hep-th/9704080.}
\lref\seib{N. Seiberg, \prl {\bf 79} 3577-3580 (1997), hep-th/9710009.}
\lref\sen{A. Sen, hep-th/9709220.}
\lref\dr{M. Dine and A. Rajaraman, hep-th/971074.}
\lref\FFI{M. Fabbrichesi, G. Ferretti and R. Inego, hep-th/9806018.}
\lref\TR{W. Taylor and M. Raamsdonk, hep-th/9806066.}
\lref\EG{R. Echols and J. Grey, hep-th/9806109.}
\lref\OY{Y. Okawa and T. Yoneha, hep-th/9806108.}
\lref\FFII{M. Fabbrichesi, G. Ferretti and R. Inego, hep-th/9806166.}
\lref\MSW{J. McCarthy, L. Susskind and A. Wilkins, hep-th/9806136.}
\lref\dos{M.R. Douglas, H. Ooguri and S. Shenker, \pl {\bf B402} 36-42
(1997), hep-th/9702203.}
\lref\do{M.R. Douglas and H. Ooguri, hep-th/9710178.}
\lref\DHN{B. de Witt, J. Hoppe and H. Nicolai, \np {\bf 305} 545 (1998).}
\lref\KT{D. Kabat and W. Taylor, hep-th/9711078, hep-th/9712185.}
\lref\motl{L.~Motl, hep-th/9701025.}
\lref\bs{T.~Banks, and N.~Seiberg, \np {\bf B497}(1997) 41-55,
hep-th/9702187.} 
\lref\dvv{R.~Dijkgraaf, E.~Verlinde and H.~Verlinde, hep-th/9703030.}
\lref\GHV{S. Giddings, F. Hacquebord and H. Verlinde, hep-th/9804121.}
\lref\BBN{G. Bonelli, L. Bonora and F. Nesti, hep-th/9805071.}
\lref\IMSY{N. Itzhaki, J. Maldacena, J. Sonnenstein and S. Yankielowicz,
hep-th/9802042.}
\lref\BBPT{K. Becker, M. Becker, J. Polchinski, A. Tseytlin, Phys. Rev. 
{\bf D56} 3174-3178 (1997), hep-th/9706072.}
\lref\CT{I. Chepelev, A.A. Tseytlin, \np {\bf B515} 73-113 (1998), 
hep-th/9709087.}
\lref\BB{K. Becker, M. Becker, \np {\bf B506} 48-60 (1997), 
hep-th/970591.}
\lref\matyt{A. Matytsin, \np {\bf B411} 805-820 (1994),
hep-th/9306077.}
\lref\DV{J.-M. Daul and V.A. Kazakov, \pl {\bf B355} 371-376 (1994), 
hep-th/9310165.}
\lref\wynt{T. Wynter, \pl {\bf B415} 349-357 (1997), hep-th/9709029.}
\lref\AdS{J. Maldacena, hep-th/9711200.}
\lref\GG{M.B. Green and M. Gutperle, \pl {\bf B398} 69-78 (1997), 
hep-th/9612127.}
 
\vskip6pt

\centerline{Thomas Wynter\footnote{$^\circ$}{{\tt
wynter@wasa.saclay.cea.fr}}
}

\centerline{{\it  Service de Physique Th\'eorique, C.E.A. - Saclay,
  F-91191 Gif-Sur-Yvette, France}} 

\vskip .3in
\baselineskip10pt{ We study the loop expansion for the low energy
effective action for matrix string theory. For long string
configurations we find 
the result depends on the ordering of limits. Taking $g_s\rightarrow
0$ before $N\rightarrow\infty$ we find free strings. Reversing the
order of limits however we find 
anomalous contributions coming from the
large $N$ limit that invalidate the loop expansion. We then embed
the classical instanton solution corresponding to a high energy string 
interaction into a long string
configuration. We find the instanton has a loop expansion 
weighted by fractional positive powers of $N$. Finally we identify the 
scaling regime for which  interacting long string configurations have a
loop expansion with a well defined large $N$ limit. The 
limit corresponds to large ``classical'' strings  and can be 
identified with the ``dual'' of the 't~Hooft limit, 
$g_{SYM}^2\sim N$. }

\bigskip
\rightline{SPhT-98/059}

\Date{June 1998 }

\baselineskip=16pt plus 2pt minus 2pt
\bigskip

\newsec{Introduction}

One of the most important technical mysteries of the matrix theory
\bsfs\ 
approach to non-perturbative description of M-theory/string theory is
the meaning of the large $N$ limit. All concrete calculations to date
have essentially been performed at finite $N$ (see \banks\BS\Tayl\
for reviews  and further references). Finite $N$ matrix theory
has been identified by Susskind\suss\ as having a physical meaning: it
corresponds to discrete light-cone quantized M theory and the proposal
has even been given a concrete ``derivation'' \seib\sen\ .  
However various
problems with this identification have emerged. In particular it has
been argued that three graviton scattering in flat space cannot be
reproduced by the finite $N$ matrix model \dr\ .  Recently a number of
papers have appeared on this issue \FFI\TR\EG\OY\FFII\ claiming and
disclaiming the correctness of finite $N$ matrix theory in this
context. 
It has also been suggested in \banks\BS\ that the large $N$ limit
might resolve any disparity, further arguments for this being given in \MSW\ . 
Furthermore it appears that in
a curved background it is impossible to describe even two body
scattering with a finite number of matrix variables \dos\do\ . 
Infinite matrices
are also required for a correct supergravity correspondence for more
general objects such as spherical membranes \DHN\KT\ . It is thus 
crucial to
understand the large $N$ limit. It is not even obvious that the
large $N$ limit is well-defined.

In this article we study the domain of validity of the loop expansion
for the low energy effective action of matrix string
theory \motl\bs\dvv\ . 
We will
focus on long string configurations interacting at one or two
points. For string interactions we take as a background the instanton
high energy string interactions constructed in \GHV\ . We start by
studying long string configurations. We find the result depends on the
ordering of limits. Taking $g_s\rightarrow 0$ before
$N\rightarrow\infty$ leads to free strings. Reversing the limits
however we find anomalous contributions from neighbouring
eigenvalues on the string worldsheet. These lead to $L$ loop
contribution to the effective action being weighted with a factor 
$N^{2(L-1)}$ indicating that the loop expansion is no longer valid. 
We then imbed the instanton two string interaction into two long
interacting strings and calculate the effective action about this
instanton background. We find that the $L$ loop contribution is 
weighted by $g_s^{2\over 3}N^{(L-{2\over 3})}$. This again indicates 
that the
loop expansion is not valid. Finally we identify a domain in which the
effective action is valid for interacting long strings. The limit 
corresponds to large classical strings of size $\sqrt{N}$, 
and can be identified with the ``dual'' to the 
`t~Hooft limit: $g_{SYM}\sim\sqrt{N}$. Curiously this limit
corresponds to the boundary limit found in \IMSY\ separating
the supergravity description from the orbifold CFT description.

\newsec{Low energy effective action for matrix string theory}

The form of the loop expansion for the effective action for matrix 
theory and its various compactifications has been studied in \BBPT\CT\ . 
Below we briefly recall the argument in the context of matrix string
theory.

We will concentrate on the higher derivative bosonic terms and for 
compactness of notation denote symbolically by $F$ both the two
dimensional gauge field $F_{\alpha\beta}$ and the derivatives of the
8 transverse coordinates  $\partial_{\alpha}X^I$. We rescale the
fields so that the loop counting
parameter, $1/g_{SYM}^2=\alpha'g_s^2$, appears as a 
multiplicative factor in front of the action (it is convenient to set
$\alpha'=1$ in what follows)~:
\eqn\lpct{
S\sim g_s^2\int\,d^2\sigma\,\tilde F^2\quad
{\rm with}\quad \tilde A=\,{A_{\alpha}\over g_s},\,
{X^I\over g_s}.}
The effective action expressed as a sum over loops is then given by
\eqn\lpef{
W\sim\sum_{L=0}^{\infty}\,\,{1\over g_s^{2(L-1)}}\,
\int\,d^2\sigma\,\tilde{\cal L}_L\bigr({X\over g_s},
{F\over g_s}\cdots\bigr).}
where $\tilde{\cal L}_L$ is the contribution to the effective action
from $L$ loops and the dots represent higher order derivative terms
Observing that $\tilde{\cal L}_L$ has the dimension of 
{\it length}$^{2(L-1)}$, and using dimensional analysis we arrive at
the effective action
\eqn\lpefo{
W\sim\int\,d^2\sigma\,F^2\,+\,
\sum_{L=1}^{\infty}{\cal L}_L
\quad{\rm with}\quad
{\cal L}_L=\sum_{n=2}^{\infty} g_s^{2n-2}
\,{F^{2n}\over X^{4n+2L-4}}.}
By $F^{2n}$ we simply mean the bosonic terms with $2n$ derivatives.

Since the two derivative term is not renormalized in this theory
the $L$ loop terms start at $F^4$ or higher. 
In fact an important element of the matrix theory conjecture is that
the $L$ loop term starts at $F^{2L+2}$. This is a necessary
requirement for matrix theory to correctly reproduce 
supergravity results. This dominant contribution should correspond to
classical supergravity. To date this has only been checked
up to the two loop level \BB\BBPT\ . 

\newsec{Long strings}

We now apply the above analysis to long string configurations of 
\motl\dvv.
We will do this explicitly for the $F^4$ contribution from the one 
loop calculation. The general form of the 
higher order contributions will then be easily discussed. 

At one loop the calculation of the effective action reduces to the
calculation of the determinant from the quadratic fluctuations of the
off-diagonal fields (bosonic, fermionic and ghost) in the background
of the diagonal elements forming the long string configuration. The
quadratic part of the Lagrangian is
\eqn\quadlag{
{\cal L}=\sum_{j=1}^{N-1}\, w^{\mu *}_j\bigl(D_j^2\eta_{\mu\nu}
-2{i\over g_s}F^j_{\mu\nu}\bigr)w^{\nu}_j
+\eta_j^{\dagger}\slash \hskip -6.5pt D_j\eta_j
+c_j^*D_j^2c,}
where all fields are defined on the interval
$\sigma\,\epsilon\,[0,2\pi N]$. 
\footnote{$^\diamond$ }{A fuller discussion of why the fields are defined on 
the interval $[0,2\pi N]$ rather than $[0,2\pi]$ can be found in 
\wynt\ .}
The fields $w$ and $\eta$ are the off-diagonal elements of the bosons and 
fermions and $c$ are the off diagonal elements
of the ghosts. The covariant derivatives $D_j$ and the
field strength $F_j$ are given by 
\eqn\defj{\eqalign{
D_j=&\partial-{i\over g_s}(a(\sigma+2\pi j)-a(\sigma))
\quad{\rm and}\quad\cr
F^j =& f(\sigma+2\pi j)-f(\sigma)
\quad{\rm with}\quad
f_{\mu\nu}(\sigma)=\partial_{\mu}a_{\nu}-\partial_{\nu}a_{\mu}.}}
The fields $w_j$ and $a$ are related to the original matrix elements
by
\eqn\wjwij{
w_j(\sigma+2\pi i)=w_{ij}(\sigma)
\quad{\rm and}\quad
a(\sigma+2\pi i)=a_i(\sigma)
\quad{\rm with}\quad
\sigma\,\epsilon\,[0,2\pi].}
Integrating over the off-diagonal elements leads to a series of terms
in powers of $F$, the lowest order being
\eqn\ffour{
\sum_{j=1}^{N-1}
\int\,d\tau\,\int_0^{2\pi N}\,d\sigma\,
{F_j^4\over X_j^7}{1\over N}\sum_p
\bigl(1+{p^2\over X_j^2 N^2}\bigr)^{-{7\over 2}}
\quad{\rm with}\quad
X_j(\sigma)=|X^I(\sigma)-X^I(\sigma+2\pi j)|.}
The sum over $p$ is the sum over discrete momenta $p/(2\pi N)$ around
the compact $\sigma$ direction. 
For our purposes the precise numerical factors and the tensor
structure of the $F^4$ term are not important.

To keep a fixed total string length in the limit
$N\rightarrow\infty$ we now rescale the coordinates 
$(\sigma,\tau)\rightarrow({1\over N}\sigma,{1\over N}\tau)$. This
leads to an overall factor of $1/N^3$ ($N^{-4}$ from the $F^4$ term,
$N^2$ from the $d^2\sigma$ and $N^{-1}$ from in front of the sum over
$p$). Naively, taking into account the sum over $j$, one would conclude
that the term \ffour\ scales away with a factor of $1/N^2$. Of course
this is false since for small $j$ (mod $N$) there is a singular
contribution. Indeed for small $j$ we have 
\eqn\smj{
F_j=j{2\pi\over N}\partial_{\sigma}f
\quad{\rm and}\quad
X_j=j{2\pi\over N}\partial_{\sigma}X}
This leads to an anomalous, $N$ independent, local contribution to the
action~: 
\eqn\anom{
\int\,d^2\sigma\,
{(\partial_{\sigma}f)^4\over(\partial_{\sigma}X)^7}
\sum_{j=1}^{\infty}{1\over j^3}\sum_{p=-\infty}^{\infty}
\bigl(1+{p^2\over j^2(\partial_{\sigma}X)^2 }\bigr)^{-{7\over 2}}.}
This type of anomaly is familiar from matrix models of 2D gravity
\matyt\DV\ . 

The factors of $N$ can be very simply deduced for the higher order
derivative terms to the one loop effective action. For the terms with
$F^{2n}$ the powers of $N$ are 2 from the
$d^2\sigma$, $4n$ from the $(\partial_{\sigma}f)^{2n}$ and $-4n-2$
from the $(\partial_{\sigma}X)^{-4n-2}$. In other words all the one
loop terms give local $N$ independent contributions to the effective
action.

An identical reasoning can be applied to the higher loop
contributions. One does not need to know the precise index structure.
Since one sums over all indices, there is guaranteed to be a part of
the sum which gives the most singular contribution. Picking out these
most singular terms one finds that the effective action \lpefo\ for
the long string configurations takes the form
\eqn\sanomal{
{\cal L}_L=N^{2(L-1)}\sum_{n=2}^{\infty}g_s^{2n-2}
\,{(\partial_{\sigma} f)^{2n}\over (\partial_{\sigma} X)^{4n+2L-4}}.}
In other words the $L$ loop contribution is weighted by a factor
$N^{2(L-1)}$. All this means is that the calculation of the effective
action in term of a perturbative loop expansion is not valid. Notice
however that if we first take the limit $g_s\rightarrow 0$ before we
take the large $N$ limit all the loop contributions disappear. Stated
more carefully, one sees that if the matrix theory hypothesis is true,
and the $L$ loops contribution to the effective action starts at
$n=L+1$, the loop contributions disappear provided $g_s\sim1/N$.

\newsec{String interactions}
Recently finite instanton like solutions to the 
classical equations of motion have been found which correspond to 
string interactions \GHV\ (see also \BBN\ ). 
A novel property of these solutions is that there
is a minimal non-zero distance between the strings. They 
split and join without touching by
stepping off into the noncommutative part of the space. Technically
the fact that they have a minimal separation means that the
fluctuations about these configurations remain massive throughout the
interaction region and leads to the hope that we might have some control over
the calculation of the effective action. 

Ultimately we are interested in the large $N$ limit so we will embed
the solution found in \GHV\ into a configuration of two long strings
joining to become a single long string. If the two strings are of the
identical length the embedding is particularly simple.

First let us recall the construction of \GHV\ to which we refer the
reader for fuller details. Instanton solutions
are found by studying the four dimensional self dual YM equations
dimensionally reduced to two dimensions~:
\eqn\sfdl{\eqalign{
F_{w,\bar{w}}=&-{i\over g_s}[X,\bar{X}]\cr
D_wX=&0\cr
D_{\bar{w}}\bar{X}=&0}}
where we have defined complex coordinates
\eqn\defX{
X={1\over\sqrt{2}}(X^1+iX^2),\,\,\bar{X}={1\over\sqrt{2}}(X^1-iX^2),}
and similar complex coordinates for the gauge fields $A$ and
$\bar{A}$. 

Single valued matrix configurations corresponding to interacting
strings can be generated by gauge transforming the diagonal
multivalued matrix with a gauge transform $U$ that creates, by Wilson
lines, the correct monodromies around the branch points \wynt\ . This
leads to delta function singularities in the field strength at the
interaction points. The key observation of \GHV\ was that these
singularities can be removed once we are working with complex
coordinates $X$, by using a complexified ``gauge'' transformation $G$
which also has a singularity at the origin tuned in such a way as to
leave a singularity free field strength. 
\subsec{$2\times 2$ matrices}
For the case of two eigenvalues we have 
\eqn\intsol{\eqalign{ X=&UG\hat{X}
G^{-1}U^{\dagger}\cr
A=&-ig_s\bigl[G^{-1}(\partial_wG)+U^{\dagger}(\partial_wU)\bigr],}} 
where the diagonal matrix $\hat{X}$, and the matrices $U$ and $G$ are
given by 
\eqn\XUG{ 
\hat{X}=B\sqrt{\bar{w}}\tau_3, 
\quad U=e^{{1\over 8}\ln{w\over\bar{w}}\tau_1} 
\quad{\rm and}\quad
G=e^{\alpha(w\bar{w})\tau_1}.}  
The unitary matrix $U$ generates the
monodromy around the branch point so that the matrix $X$ remains
single-valued even though its eigenvalues interchange.  This ansatz
automatically satisfies the last two equations of \sfdl\ with the
first equation leading to a differential equation for $\alpha$ 
\eqn\difa{ 
(\partial_r^2+{1\over r}\partial_r)\alpha= 
{8B^2\over g_s^2}r\,\sinh {2\alpha} 
\quad{\rm with}\quad
\alpha\rightarrow\cases{0\quad{\rm for}\quad r\rightarrow\infty\cr
-{1\over 4}\ln{r}\quad{\rm for}\quad r\rightarrow 0} } 
where
$r=\sqrt{w\bar{w}}$ is the radial distance from the branch point.  The
boundary conditions are necessary for a finite solution. In
particular the second boundary condition ensures that there are no
${1\over w}$ pole terms in the gauge field $A$ and hence no delta
function singularity in the field strength $F_{w\bar{w}}$.

The differential equation can be given a dimensionless form by
absorbing the coupling constants into a redefinition of the radial
coordinate so that 
\eqn\alph{
\alpha(r,B,g_s)=\alpha(r')\quad{\rm with}\quad
r=\biggl({g_s^2\over 8B^2}\biggr)^{1\over 3}r'.}
Before turning to the effective action let us give here the expressions
for $X$ and $F_{w\bar{w}}$~:
\eqn\XF{\eqalign{
&X=B(x_3\tau_3+ix_2\tau_2)\cr
&F_{w\bar{w}}={2iB^2\over g_s}f_1\tau_1}
\quad\quad{\rm where}\quad\quad\eqalign{
x_3=&\sqrt{\bar{w}}\,\cosh{\alpha},
\quad
x_2=\sqrt{w}\,\sinh{\alpha}\cr
f_1=&r\,\sinh{2\alpha}.}}
For simplicity we do not include the final gauge transformation $U$
\XUG\ .
We now focus on the $F^4$ contribution from the
one loop calculation, and use a formalism that generalizes easily for
large $N$. The quadratic part of the fluctuation
Lagrangian reads
\eqn\flucL{
{\cal L}=\Tr\bigl[ V^{\mu }\bigl(D^2\eta_{\mu\nu}
-2{i\over g_s}F^{\mu\nu}\bigr)V^{\nu}
+\eta^{\dagger}\slash \hskip -6.5pt D\eta
+c^*D^2c\bigr],}
where $V$ and $\eta$ are the bosonic and fermionic fluctuations and 
$c$ are the ghost fluctuations. All background fluctuations couple to
the background fields and are massive. It is convenient to express the 
mass term for $V^{\mu}$ and the quadratic coupling to the background 
$F^{\mu\nu}$ in terms of the $SU(2)$ generators. The mass term comes
from the double commutator $[X,[\bar{X},V]]+{\rm c.c.}$ in the kinetic
$D^2$ term of \flucL\ .
\eqn\VF{\eqalign{
{\cal L}_{\rm mass}=&{8B^2\over g_s^2}
\bigl[|v_1|^2(x_3^2+x_2^2)+|v_2|^2x_3^2
+|v_3|^2x_2^2\bigr]\cr
{\cal L}_{\rm F}=&{8B^2\over g_s^2}
\bigl[v_2^0v_3^9-v_2^9v_3^0\,
-\,v_2^1v_3^2+v_2^2v_3^1\bigr]\,f_1,
}}
where lower indices are group indices and upper indices correspond
to spacetime indices. The indices $0$ and $9$ correspond to the two
dimensional gauge fields. We have written the terms in this form so
that they generalize easily to the case of strings of length $N$.
The $F^4$ contribution is (up to numerical factors) given by~:
\eqn\Finst{
{\cal S}_{F^4}\,=\,{8B^2\over g_s^2}\int\,d^2\sigma\,h(x_2,x_3,f_1)
=\int\,d^2\sigma'\,\tilde{h}(r',\alpha(r'))}
where
\eqn\defh{\eqalign{
h(x_2,x_3,f_1)=&f_1^4\biggl[
{x_3^2+x_2^2\over x_3^2 x_2^2(x_3^2-x_2^2)}
-2{\ln{x_3^2}-\ln{x_2^2}\over (x_3^2-x_2^2)^3}
\biggr]\cr
=&r\,[2\sinh^2{2\alpha}\cosh{2\alpha}-\sinh^4{2\alpha}\ln{\coth{\alpha}}]\cr
=&\bigl({g_s^2\over 8B^2}\bigr)^{1\over 3}
\tilde{h}(r',\alpha(r')).}}
The precise form of $h(x_2,x_3,f_1)$ is not important for our
purposes. All that is important is that it consists of a factor of $r$
multiplied by a function depending only on $\alpha$ so that on rescaling
the coordinates according to \alph\ we arrive at the last line of \defh\ 
and the right-hand side of \Finst\ .

A simplifying assumption used in the above is that the size of the
instanton is much less than the size of the compact direction
$\sigma$. It is then a good approximation to use the 
instanton ansatz \XUG\ , which assumes the space is non-compact. The
instanton is then assumed to be glued into a globally defined configuration
corresponding to a branched covering of the cylinder \wynt\ with the
instanton sitting at the branch point. 

We see that the result \Finst\ is independent of $g_s$ and $B$. 
This reasoning can be generalized to show that all the terms in the
1-loop expansion are independent of $g_s$ and $B$. 
Each factor of $F$ in \lpefo\ contributes $B^2r/g_s$ and each factor of 
$X^2$ contributes $B^2r$. The net result being that all terms in the 
1 loop contribution can be written as
\eqn\instone{
{B^2\over g_s^2}\int\,d^2\sigma\,r\,f(\alpha),}
where $f$ is a function of $\alpha$ alone. After rescaling of the
coordinates \alph\ such contributions are independent of $g_s$ and $B$.
Applying this reasoning to the higher loop terms in \lpefo\ it is
easy to see that the $L$ loop contribution is weighted with a factor
of $(g_sB^2)^{-{2\over 3}(L-1)}$.

\subsec{$2N\times 2N$ matrices}
Let us now imbed this solution, which consists of just two eigenvalues,
into a configuration of two long strings of equal length $N$ joining into
a single string. This will permit us to relax the condition on the
size of the instanton to be less than the total
length of the long string rather than less than the length of an
individual string. This is necessary physically since the parameter
$B$ of the instanton solution is proportional to $1/\sqrt{N}$ and
hence leads via \alph\ to an instanton of size $N^{1\over 3}$,
i.e. physically the instanton will spread out over 
{\cal O}$(\root 3 \of N)$
individual strings.

To do this we simply tensor the $2\times 2$ solution by diagonal
$N\times N$ blocks, with the diagonal elements forming cycles of
length $N$. The complex coordinates $w$ and $\bar{w}$ then sit on the
cylinder of length $2\pi N$. The matrices for the instanton
configuration $X$ and $F_{w\bar{w}}$ and the matrix for the background
fluctuations $V$ are given by
\eqn\XFN{\eqalign{
X=&B\bigl(x_3\otimes\tau_3+ix_2\otimes\tau_2\bigr)\cr
F_{w\bar{w}}=&{2iB^2\over g_s}f_1\otimes\tau_1\cr
V=&v_1\otimes\tau_1+v_2\otimes\tau_2+v_3\otimes\tau_3}.}
The matrices $x_2$, $x_3$ and $f_1$ are diagonal matrices with
the entries forming cycles of length $N$, i.e.
\eqn\xxf{\eqalign{
(x_3)_{ij}=&\delta_{ij}\,x_3(\sigma+2\pi(i-1),\tau)\cr
(x_2)_{ij}=&\delta_{ij}\,x_2(\sigma+2\pi(i-1),\tau)\cr
(f_1)_{ij}=&\delta_{ij}\,f_1(\sigma+2\pi(i-1),\tau),}
\quad{\rm with}\quad
\eqalign{
x_3(\sigma,\tau)=\sqrt{\bar{w}(\sigma,\tau)}\,\cosh{\alpha(\sigma,\tau)}\cr
x_2(\sigma,\tau)=\sqrt{\bar{w}(\sigma,\tau)}\,\sinh{\alpha(\sigma,\tau)}\cr
f_1(\sigma,\tau)=r(\sigma,\tau)\,\sinh{2\alpha(\sigma,\tau)},}}
This corresponds to long strings of length $N$. The $SU(2)$ structure
of the instanton then splits and joins the two long strings. It is trivial
to see that the differential equation \difa\ remains unchanged.

The matrices $V$ are general hermitean $N\times N$ matrices. 
They also have a periodicity condition, since they end and start on the
diagonal matrix background, provided by \xxf .
\eqn\vp{
v_{ij}=v_i(\sigma+2\pi(j-i))}
The calculation of the effective action is a simple generalization of
that for $N=1$. The key difference being that whereas before the
commutators give rise to products of scalars $x_3$, $x_2$, $f_1$ they
now give rise to anticommutators for the $N\times N$ matrices $x_3$,
$x_2$, $f_1$. For example the contribution to the $v_2$ mass term
is given by
\eqn\massv{\eqalign{
\int_0^{2\pi}d\sigma
{1\over 2}\Tr\bigl[v_2\otimes\tau_2[X,[\bar{X},v_2\otimes\tau_2]]+{\rm
c.c.}\bigr]
=&2\int\Tr[v_2\{x_3,\{\bar{x}_3,v_2\}\}+{\rm c.c.}]\cr
=&4\int\sum_{i,j}\bigl|(v_2)_{ij}\bigr|^2\bigl|(x_3)_i+(x_3)_j\bigr|^2\cr
=&8\int_0^{2\pi N}\hskip -3pt d\sigma
\sum_j\bigl|(v_2)_j\bigr|^2
\bigl|(x_3)_j\bigr|^2.}}
where 
\eqn\defx{
(x_3)_j={1\over 2}\bigl(x_3(\sigma,\tau)+x_3(\sigma+2\pi
j,\tau)\bigr).}
The total mass term and interaction vertex read
\eqn\VFN{\eqalign{
{\cal L}_{\rm mass}=&{8B^2\over g_s^2}
\,\sum_j\bigl(\bigl|(v_1)_j\bigr|^2(\,(x_3)_j^2+(x_2)_j^2\,)
+\bigl|(v_2)_j\bigr|^2\,(x_3)_j^2
+\bigl|(v_3)_j\bigr|^2\,(x_2)_j^2\cr
{\cal L}_{\rm F}=&{8B^2\over g_s^2}
\,\sum_j\bigl[(v_2^0)_j(v_3^9)^*_j-(v_2^9)_j(v_3^0)^*_j\,
-\,(v_2^1)^*_j(v_3^2)_j+(v_2^2)_j(v_3^1)^*_j\bigr]\,(f_1)_j,}}
where
\eqn\deff{
(f_1)_j={1\over 2}\bigl(f_1(\sigma,\tau)+f_1(\sigma+2\pi
j,\tau)\bigr).}
These are simple generalizations of equations \VF\ where for each $j$
one replaces the instanton background fields $x$ and $f$ by their
average over points separated by a distance $2\pi j$ in the $\sigma$
direction. 

The effective action likewise generalizes straightforwardly. For the
one loop $F^4$ contribution for example one sums
over $j$ the result \Finst\ with the masses and fields strengths in
\defh\ replaced by their average values over points separated by 
$2\pi j$.

We are now in a position to analyse the behaviour at large $N$ of the
effective action for the instanton configuration. The crucial
difference between the large $N$ behaviour of the long string
effective action and the part of the effective action for the
instanton is that for the 
instanton there is no large $N$ anomaly. This is due to the fact that
one takes the average over background field values separated by $2\pi
j$ in the $\sigma$ direction, not the difference. There is thus no
singular behaviour for small $j$ and if the size of the instanton is
such that it spreads out over a large number of strips, the sums over
$j$ can be replaced by integrals. The $F^4$ term in the one loop
expansion is thus given by
\eqn\linst{\eqalign{
{\cal S}_{F^4}
=&{8B^2\over g_s^2}\int d\tau d\sigma\sum_j
h((x_2)_j,(x_3)_j,(f_1)_j)\cr
=&{8B^2\over g_s^2}\int d\tau d\sigma_1 d\sigma_2 
h((x_2),(x_3),(f_1))\cr
=&\bigl({g_s^2\over B^2}\bigr)^{1\over 3}
\int d\tau' d\sigma_1' d\sigma_2'
\tilde{h}(r_1',\alpha(\sigma_1',\tau_1'),r_2',\alpha(\sigma_2',\tau_2'))}}
where in the second line we have defined
\eqn\xxp{
(x_3)={1\over 2}\bigl(x_3(\sigma_1,\tau)+x_3(\sigma_2,\tau)\bigr),}
and in the final line we have rescaled all coordinates to the natural
size of the instanton, using \alph\ and we have also used the last
line of \defh\ .

Finally the $N$ dependence of the $F^4$ contribution is determined by
the $N$ dependence of $B$ which specifies the asymptotic behaviour
of the instanton/branch point. For the asymptotic behaviour to have
a physically sensible large $N$ limit we see from equation \XF\ and the
fact that we rescale coordinates $w\rightarrow\sqrt{N}w$ that the
constant $B$ must scale as $B\sim 1/\sqrt{N}$. The $F^4$ term thus
scales as $g_s^{2/3}N^{1/3}$.

The factors of $g_s$ and $N$ for the higher derivative contributions
to the the 1 loop effective action can likewise be easily
determined. The difference from the scaling arguments given for the
case $N=1$ at the end of section(4.1) is that there is now an extra
$\int\,d\sigma$ integral from the sum over $j$ and hence an extra 
$(g_s^2/B^2)^{1/3}$ factor. All terms in the 1 loop effective action
are thus seen to scale as $g_s^{2/3}N^{1/3}$.

For the higher loop terms in the effective action the dominant
contribution comes from planar diagrams. In terms of the matrix theory
conjecture the first such planar diagram at $L$ loops is hoped to be
the term $F^{2L+2}$. For our purposes each planar loop brings an extra index
and hence an extra $\int\,d\sigma$ integral in addition to the 
$\int\,d^2\sigma$ integral of \lpefo\ . This leads to the $L$ loop
weight
\eqn\lscl{
S_{L}\sim g_s^{2\over 3}B^{-2(L-{2\over 3})} \sim 
g_s^{2\over 3}N^{(L-{2\over 3})}.}
As for the long string configurations into which the instanton is
embedded we conclude that the loop expansion is not valid.

\newsec{Large $N$ limits}
In this section we search for a large $N$ limit in which the loop
expansion is well defined both for the long strings and for their
interactions (instanton configurations). In other words we allow both
the string coupling constant and the size of the string to scale with
powers of $N$~:
\eqn\Xgscal{\eqalign{
X,F\sim N^x&\Rightarrow B\sim N^{x-{1\over 2}}\cr
g_s\sim N^g.}}
We then look for the region in $x$, $g$ space where all terms in the
loop expansions scale with non-positive powers of $N$. 

From the anlysis of the long string
congfigurations we have~:
\eqn\xglsconstr{
L-1+(n-1)g-(n+L-2)x\leq 0}
If we further assume that $n\geq L+1$ as is required for the matrix
theory to be correct the inequality leads to the two inequalities
\eqn\xglsconsrtt{
g\leq x
\quad\quad{\rm and}\quad\quad
g\leq 2x-1.}
The analysis of the loop expansion for the instanton/string
interaction consists of two cases depending on whether or not the
world sheet scale of the instanton \alph\ spreads out over many
strips. If $g>x-1/2$ the instanton size will scale as a positive power
of $N$, and the analysis of section 4.2 is then appropriate. Imposing,
in this case, that all terms in the loop expansion scale with non
positive powers of $N$ leads to the inequality
\eqn\xginconstr{
g\leq(3L-2)(x-{1\over 2})
\quad{\rm which\,\,implies}\quad g\leq x-{1\over2}
\quad{\rm and}\quad 
x\geq{1\over2},}
in contradiction with the domain of validity of \xginconstr\
itself. We are left with the case $g\leq x-1/2$, in which the size of
the instanton is less than an individual strip, and for which the
analysis of section 4.1 is appropriate. We thus have the inequality
\eqn\gxineq{
g\geq1-2x.}
We plot the inequalities \xglsconsrtt\gxineq\ in the graphic below
\vskip 20pt
\hskip 100pt
\epsfbox{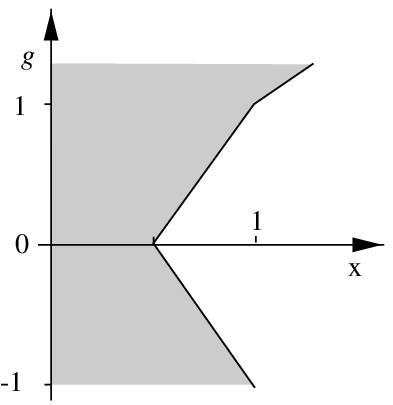}
\vskip 5pt
\centerline{Fig. 1. $x$, $g$ parameter space for which the loop}
\centerline{expansion is valid (shown in white)}
\vskip 10pt
For most of the available parameter space all of the
terms in the loop expansions are scaled away. At the point $x=1/2$,
$g=0$, however, all terms in the loop expansion for the instanton
contribute. This limit corresponds to large classical strings of size
$\sqrt{N}$ in $\alpha'$ units with $\sqrt{N}$ individual diagonal
elements/strips of the world sheet occupying an interval of length 
$\sqrt{\alpha'}$. This resembles the scaling limits studied
in the AdS - SYM correspondence (see \AdS\ and references thereto). 
Indeed an alternative way of thinking
of this scaling regime is as a fixed string size but a rescaling 
of $\alpha'$ by $1/N$, which via
the relation $g_{SYM}^2=1/g_s^2\alpha'$ leads to $g_{SYM}^2\sim
N$. This is the ``dual'' of the 't Hooft limit. Interestingly it is
also precisely the point found in \IMSY\ separating the CFT
orbifold description from the SUGRA description.

\newsec{Conclusions}
We have seen in the previous sections that the loop expansions for the
case of physical interest, that of interacting matrix strings, is ill
defined in the limit $N\rightarrow\infty$. This does not imply that the
theory itself is ill defined, it just
highlights the fact that it is not justified to use such an expansion.
To decide whether the theory does or does not have a well defined 
large $N$ limit 
would require finding some way of integrating out, at least partially, 
some of the non-perturbative contributions. It could 
be that in the full non-perturbative calculation the ``mass'' of
the off-diagonal elements connecting neighbouring strips is smoothed
out in such a way that all the large $N$ anomalies found in section(3)
disappear. A hint that this might be the case comes from the study of
the effective action for the diagonal elements in the $0$ dimensional
matrix model. For $N=2$ it is possible to integrate out the
off-diagonal elements explicitly and one finds that the singularity for
coinciding diagonal elements is resolved by the appearance of extra
massless fields \GG\ .

We have found that there is a non-trivial scaling limit in
which the loop expansion is well defined in the $N\rightarrow\infty$. 
It corresponds to the ``dual'' of the 't Hooft limit, $g_{SYM}^2\sim N$.

Finally in the calculation of the fluctuations about the instanton
careful attention should be paid to the translation and scale
modes. These are not important for our purposes since we are focusing
on the large $N$ limit, but could well contain important contributions
to the full result. We leave the investigation of this point to future
work.

\newsec{Acknowledgements} I would especially like to thank Ivan Kostov
for many fruitful discussions. I am also grateful to Esko
Keski-Vakkuri for some useful comments. Finally I thank the CERN
theory division for their hospitality where a large part of this work
was carried out.

\listrefs

\bye